\documentclass[twocolumn,tighten]{revtex4}

\usepackage{epsfig}
\usepackage{natbib}
\usepackage{ulem}
\usepackage{xspace}
\usepackage{appendix}
\usepackage{graphicx}
\usepackage{amsmath}
\usepackage{amssymb}
\usepackage[dvipsnames]{xcolor}

\begin{document}

\title{Approaching the perfect diode limit through a nonlinear interface 
}
\author{Lucianno Defaveri$^{1}$} 
\author{Alexandre A. A. Almeida$^2$}
\author{Celia Anteneodo$^{2,3}$}
\affiliation{$^1$Department of Physics, Bar Ilan University, Ramat-Gan 52900, Israel}
\affiliation{$^2$Department of Physics, PUC-Rio, Rio de Janeiro, 22453-900 RJ, Brazil}
\affiliation{$^3$Institute of Science and Technology for Complex Systems, INCT-CS, Rio de Janeiro, Brazil}

\begin{abstract}

We consider a system formed by two different segments of particles, 
coupled to thermal baths, one at each end, modeled by Langevin thermostats.
The particles in each segment interact harmonically and are subject to an on-site potential, for which, 
three different types are considered, namely, 
harmonic, $\phi^4$, and Frenkel-Kontorova. 
The two segments are nonlinearly coupled, between interfacial particles, by means of  a power-law potential, with exponent $\mu$, 
 which we vary, scanning from subharmonic to superharmonic potentials, up to the infinite-square-well limit ($\mu\to\infty$). 
Thermal rectification is investigated by integrating the equations of motion and computing the heat fluxes. 
As a measure of rectification, we use the difference of the currents resulting from baths inversion, divided by their average.  
We find that rectification can be optimized by a given value of $\mu$ that depends on the bath temperatures and details of the chains. But, regardless of the type of on-site potential considered,  the interfacial potential that produces maximal rectification approaches the infinite-square-well ($\mu\to\infty$), when reducing the average temperature of the baths.     
Our analysis of thermal rectification focuses on this regime, for which we complement numerical results with heuristic considerations. 
\end{abstract}

\maketitle

\section{Introduction}
 
A thermal diode is a material whose thermal conductivity along a given axis changes depending on the direction of the heat flux. 
After a period of little activity in the area since the pioneering works~\cite{starr36,review}, the interest in thermal rectification was boosted~\cite{review}, both theoretically ~\cite{terraneo2002,casati-diode,efficiency1,efficiency2, efficiency3} and experimentally~\cite{changSolidstate2006,kobayashiOxideThermal2009,tianNovelSolidstate2012,martinez-perezRectificationElectronic2015,tsoSolidstateThermal2016,shresthaDualmodeSolidstate2020,kasprzakHighTemperature2020}, focusing mainly on the conditions to improve the efficiency of thermal rectification. 

Asymmetry and temperature dependence of the conductivity are required to break the symmetry of inversion in flux direction~(see for instance refs. \cite{Tdependence,defaveri2021,muga2021,muga2023}). 
Asymmetry can be achieved in several ways, for instance, by mass-graded chains~\cite{efficiency3,bastida}, homogeneous chains with asymmetric interactions~\cite{benenti}, simply by the presence of impurities or defects~\cite{muga2017,alexander2020}, or coupling two or more different segments~\cite{casati-diode}. 
Temperature dependence of the conductivity can be intrinsic~\cite{muga2021} or produced by anharmonic interactions between particles or with the substrate~\cite{defaveri2021}.

On the other hand, the interface can play a crucial role in thermal conduction, as observed experimentally~\cite{strain1,strain2} and 
in simulations~\cite{interface,BaowenLi2005,BaowenLi2011,Hu2006,muga2017,defaveri2021}.   
%
The interface can affect the overlap of the phonon bands of each segment, which selects the conducting modes. 
In the case of two segments, 
the phonon bands of each decoupled segment are different due to the asymmetry or distinct characteristics of each segment, and the nonlinear interactions make the spectra temperature-dependent. Then, the band overlap  can change under flux inversion, giving rise to a preferential direction for heat conduction. 
If the coupling is weak and linear, 
we expect that similar overlaps to those obtained for isolated segments will hold for the coupled chain. As the communication between segments (controlled by the interfacial stiffness) increases, band overlap will be affected, and differences related to temperature inversion blurred~\cite{casati-diode}, 
reducing rectification. 
However, 
if the coupling were nonlinear, band overlap might become more complex~\cite{hanggi},  with strong interference of bands even for weak coupling strength, and efficiency optimization may appear under certain nontrivial conditions.

\begin{figure}[b!]
    \includegraphics[width=0.4\textwidth]{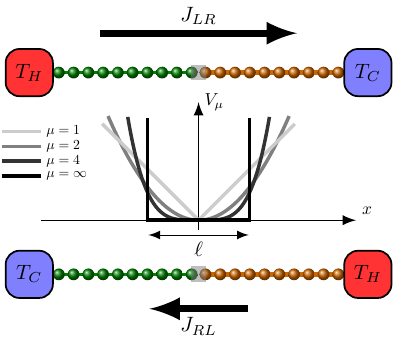}
	\caption{Schematic representation of the system formed by two different segments nonlinearly coupled and subject to thermal baths at the ends, producing rectification under flux inversion. }
	\label{fig:pic}
\end{figure}

In this context, we address the following questions. Which is the effect of a nonlinear interface in two-segment systems subject to thermal baths at the ends?
Under which conditions can this nonlinearity improve the performance of the thermal diode? 
To find insights about the answers to these questions, we consider two-segment chains of particles that are harmonically coupled and subject to a local (on-site) potential  (either harmonic,  $\phi^4$ or  Frenkel-Kontorova), while the segments are coupled through a power-law potential $V_\mu$, with exponent $\mu$, between interfacial particles. Additionally, we consider  Langevin baths connected at the ends and fixed boundary conditions.  
A schematic representation of the studied family of systems is given in Fig.~\ref{fig:pic}.

From the integration of the equations of motion, we obtain the heat currents and a quantifier of rectification, and explore the effects of the nonlinear junction. 
The paper is organized as follows. 
We define the model in Sec.~\ref{sec:model} and the methods in Sec.~\ref{sec:methods}. 
Results of numerical simulations and theoretical considerations, when varying the interfacial nonlinearity, with focus on the infinite square-well limit at low temperatures, are presented in Sec.~\ref{sec:results} and in the
Appendix. 
A discussion and concluding remarks are presented in Sec.~\ref{sec:final}.

\section{Mathematical model}
\label{sec:model}

In each of two, left ($L$) and right ($R$), segments, particles of identical mass $m$ interact harmonically with their first neighbors, with stiffness constant  $k_{L/R}$, and  are subject to an  on-site potential $V_{L/R}$, representing the interaction with a substrate of negligible thermal conductivity.
The two segments are coupled through a power-law potential. Namely, the full system is governed by  the Hamiltonian 
\begin{eqnarray} \nonumber
\mathcal{H} &=& \sum_{n=1}^{N} \frac{1 }{2 } \frac{p_{n}^{2} }{m}
+\sum_{n=1}^{M} \frac{k_{L}}{2}(x_{n}-x_{n-1}-a)^{2}
+ \sum_{n=1}^{M} V_{L}(x_n)\\ \nonumber
&~& ~ ~~ ~ +   V_\mu(x_{M+1} - x_{M}-a)   \nonumber \\
&+& \sum_{n>M}^{N} \frac{k_{R}}{2}(x_{n+1}-x_{n}-a)^{2}
+\sum_{n>M}^{N} V_{R}(x_n),
\label{eq:hamiltonianl}
\end{eqnarray}
where   $x_n$ (with $n=1,\ldots,N=2M$,  indexed from left to right) is the coordinate  of particle $n$,  $p_n$ its momentum, $a$ is the lattice constant, and 
$k_{L/R}$ are stiffness constants.  
As on-site potentials 
$V_{R/L}$, we will consider diverse cases defined below. 
Moreover, boundary conditions remain fixed, such that 
 $ x_{0} = 0$ and $x_{N+1} = (N+1)a$,  at all time. 
The end particles 1 and $N$, are immersed in 
thermal baths represented by Langevin thermostats, 
at temperatures $T_{L}$ and $T_{R}$.
Then, the equations of motion are given by
\begin{eqnarray} \label{eq:motion1}
\dot{p}_1 = m \ddot{x}_1 &=& - \frac{\partial \mathcal{H}}{\partial x_1} - \gamma \dot{x}_1 + \eta_1(t), \\ \label{eq:motionnn}
\dot{p}_n = m \ddot{x}_n &=& - \frac{\partial \mathcal{H}}{\partial x_n} \;\;\;\; \mbox{for $n = 2 \dots   N-1$},\\ \label{eq:motionN}
\dot{p}_N = m \ddot{x}_N &=& - \frac{\partial \mathcal{H}}{\partial x_N} - \gamma \dot{x}_N + \eta_N(t),
\end{eqnarray}
where $\eta_1$ and $\eta_N$ are uncorrelated stochastic Gaussian white noises with zero mean and
\begin{eqnarray}
\left\langle \eta_1(t) \eta_1(t') \right\rangle &=& 2\gamma k_B T_L \, \delta(t-t'), \nonumber \\
\left\langle \eta_N(t) \eta_N(t') \right\rangle &=& 2 \gamma k_B T_R \, \delta(t-t') \, .
\end{eqnarray}
Finally, 
in order to study the impact of the nonlinearity of the  interfacial interaction, we consider that
 \begin{eqnarray}
    V_\mu(x) = \frac{k_\mu}{\mu} \left( \frac{|x|}{\ell} \right)^\mu
     \label{eq:Vmu}
 \end{eqnarray}
 where  $k_\mu$ is a positive constant, $\ell$ is the characteristic length of the interaction and the exponent is $\mu\ge 1$ (to avoid a divergent force at the origin). 
 The exponent  $\mu\neq 2$, not necessarily integer, characterizes a restoring force 
that can gives rise to oscillations  beyond the simple harmonic case, associated to nonlinear responses~\cite{defaveri2019,dhar2019,schmelcher2018,meiners2000}. 
It encompasses two classes of nonlinearity: for $\mu > 2$ ($\mu  < 2$) the force depends super(sub)-linearly on the deformation or displacement from the equilibrium position. 
$V_\mu$ ranges from the  triangular ($\mu = 1$) to the infinite square well ($\mu \to \infty$) potentials, including the harmonic and quartic potentials. 
Even though $\ell$ could be absorbed into $k_\mu$, we keep track of both parameters separately so that in the limit $\mu \to \infty$, which is equivalent to an infinite square well, we can control the size of the well through $\ell$.

With regard to the on-site potential, we analyze different cases, starting with the harmonic pinning  given by
\begin{equation}
V(x_{n}) = \frac{A}{2 }\left(x_{n} - na\right)^{2} .
\label{eq:harmonic}
\end{equation}
In such case, the Fourier law does not hold, a flat bulk temperature profile and conductivity that scales with $N$ is known to occur due to momentum conservation~\cite{NarayanRamaswamy2002,ReviewLepriLiviPoliti2003,ReviewDhar2008}. 

As substrates that break momentum-conservation and allow 
Fourier-law, we consider the following ones.

(i) A particular case of the $\phi^4$  potential, namely  
\begin{equation}
V(x_{n}) = \frac{A}{4 }\left(x_{n} - na\right)^{4} ,
\label{eq:phi4}
\end{equation}
where $A$ is a positive constant. 
 This unbounded potential has been thoroughly studied in the literature of thermal conduction ~\cite{aokiEnergyTransport2006, savinHeatConduction2003, huHeatConduction2000, ReviewDhar2008, lefevereNormalHeat2006, liParameterdependentThermal2007}.
 
(ii) The second case  is a paradigm of periodic bounded potential, the Frenkel-Kontorova (FK) potential~\cite{huHeatConduction1998, Gillan1985i}, with period $ a$,   given by
\begin{equation}
V(x_{n}) =
\frac{A}{2\pi}
\left[1 - 
\cos \left(  2\pi x_{n}/a \right)\right].
\end{equation}
This model has also been studied as paradigm of normal heat transport~\cite{savinHeatConduction2003, huHeatConduction2000, Gillan1985i,Gillan1985ii,Gillan1985iii}. 
%

\section{Methods}
\label{sec:methods}

The equations of motion (\ref{eq:motion1})-(\ref{eq:motionN}) were numerically integrated by means of an adaptive Euler-Maruyama method ($N=2$), with a maximum time step $\Delta t = 10^{-4}$ or adaptive stochastic higher-order Runge-Kutta method ($N=20$) with maximum time step $\Delta t = 10^{-3}$~\cite{Penland2006}.
Typically, the total simulation time is $t_{max}=10^5$ for a total of $10^3$ samples. 
In all cases, 
each ($R/L$) segment was characterized by values 
 of the force amplitudes with the relation $A_R/A_L=5$, and
 $k_R/k_L=5$, in order to produce significant asymmetry. 
 That is, without loss of generality, the right-side potential was chosen to have a larger amplitude than the left-side one.   %
 In numerical experiments,  the parameters $A_L$, $k_L$, $m$, $\gamma$ and $a$ have been assigned a unit value.  
 For simplicity, we also set $k_B = 1$ and treat temperature in energy units.   
 
In the extreme $\mu\to\infty$, the interface potential given by Eq.~(\ref{eq:Vmu}), tends to an infinite square well, null for 
$x\in[-\ell,\ell]$, infinite otherwise. 
We measured the long-time value of the heat current as the average power (work per unit time) exchanged between the interfacial particles as 
\begin{eqnarray}
\label{eq:J}
J &=& \left\langle   \frac{\dot{x}_M + \dot{x}_{M+1}}{2}   \;V'_\mu ( 
 x_{M+1} - x_{M}-a ) \right\rangle 
\end{eqnarray}
where $V'_\mu(x)=dV_\mu/dx$, 
averaged over time and over realizations, typically $\mathcal{N}=10^3$ samples, unless differently specified. 
Let us note that, with the definition of Eq.~(\ref{eq:J}), we use the convention that a positive value of $J$ means that the heat flows from left to right.

As a measure of rectification, we considered the  relative difference 
\begin{equation}
R= \frac{\left|  J_{LR}   -   J_{RL}   \right|}{\left| J_{m} \right|}\;,
\label{eq:R}
\end{equation}
where $ J_{LR}  $ and $ J_{RL} $ are the absolute values of the currents in each direction (obtained by inverting the heat baths), and $J_m=(J_{LR}  +  J_{RL} )/2$. Depending on the on-site and interface potentials, either $ J_{LR}  $ or $ J_{RL} $ turns out to be the largest one~\cite{defaveri2021}.

Temperatures were characterized by the average  
$ T_{m}=(T_{L}+T_{R})/2 $, and the relative temperature difference, $ \Delta_{rel} = (T_{L}-T_{R})/T_{m} $, hence $
T_{L/R} = T_{m} (1\pm  \Delta_{rel}/2)$, and $|\Delta_{rel}|\le 2$.

\section{Results}
\label{sec:results}

\begin{figure}[h!]
    \centering
    \includegraphics[width = 0.45\textwidth]{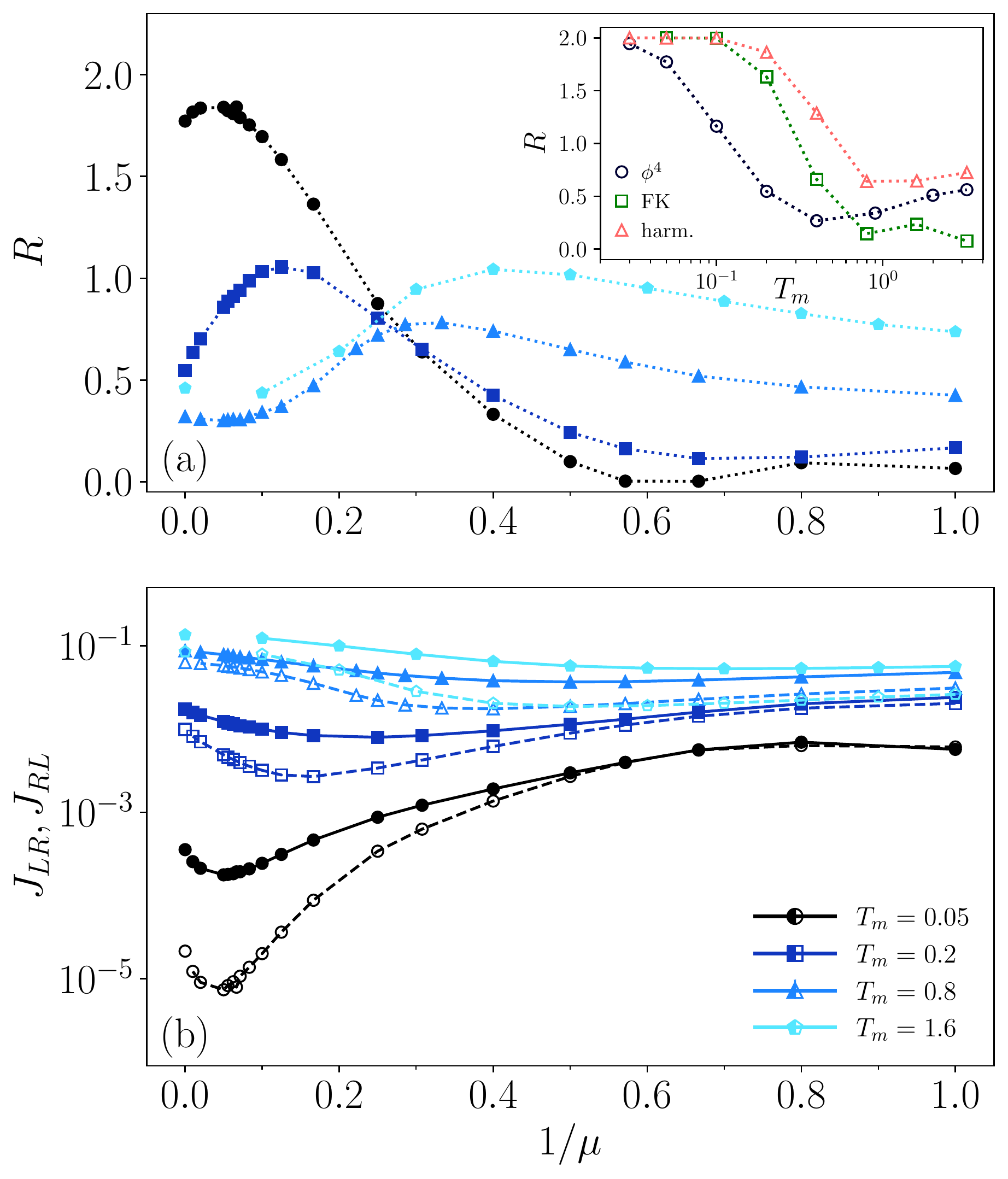}
    \caption{
    (a) Rectification factor $R$ and (b) corresponding heat currents $J_{LR}$ and $J_{RL}$, vs. $1/\mu$, for the $\phi^4$ on-site potential, and  different values of  $T_m$.   
Lines are a guide to the eye. 
$J_{LR}$ and $J_{RL}$ are plotted by filled and hollow symbols, joined by solid and dashed lines, respectively, and correspond to the time average over $10^3$ samples. 
We include the outcomes for the infinite-square-well ($1/\mu=0$), obtained through a different integration algorithm.
For this limit, in the inset of panel (a), 
$R$ is plotted vs. $T_m$, for all the on-site potentials considered. In all cases, $k_\mu = 0.5$, $\Delta_{rel}= 1.5$, and $N = 20$.
}
    \label{fig:R20-fi4}
\end{figure}

Fig.~\ref{fig:R20-fi4}(a) provides a  picture of the behavior of the rectification factor $R$ as a function of $1/\mu$ (the inverse of the exponent that characterizes the interface potential),  using a system of $N=20$ particles subject to the $\phi^4$ on-site potential. 
Each curve corresponds to a fixed average temperature $T_m$ indicated in the legend. The respective currents are also presented in Fig.~\ref{fig:R20-fi4}(b).

For each chosen value of $T_m$, the plot of $R$ as a function of $1/\mu$ follows a smooth curve that presents a single maximal value. 
 Diminishing enough the temperature $T_m$, the maximum shifts towards smaller $1/\mu$ (larger $\mu$) and approaches the perfect diode limit $R=2$.  
For the infinite square well, $R$ tends to its maximal value, as $T_m$ decreases, as shown in the inset of Fig.~\ref{fig:R20-fi4}(a). 
Such trends are common to the other two types of on-site potentials considered (as can be seen in the inset of  Fig.~\ref{fig:R20-fi4}(a), although peculiar features occur in each case at higher temperatures (see Appendix for more details).  
This finding motivates us to have a closer look at the infinite-square well potential.

 
\subsection*{Infinite-square-well interface}
\label{sec:infinite}

\begin{figure}[t!]
	\centering
 \includegraphics[width=0.455\textwidth]{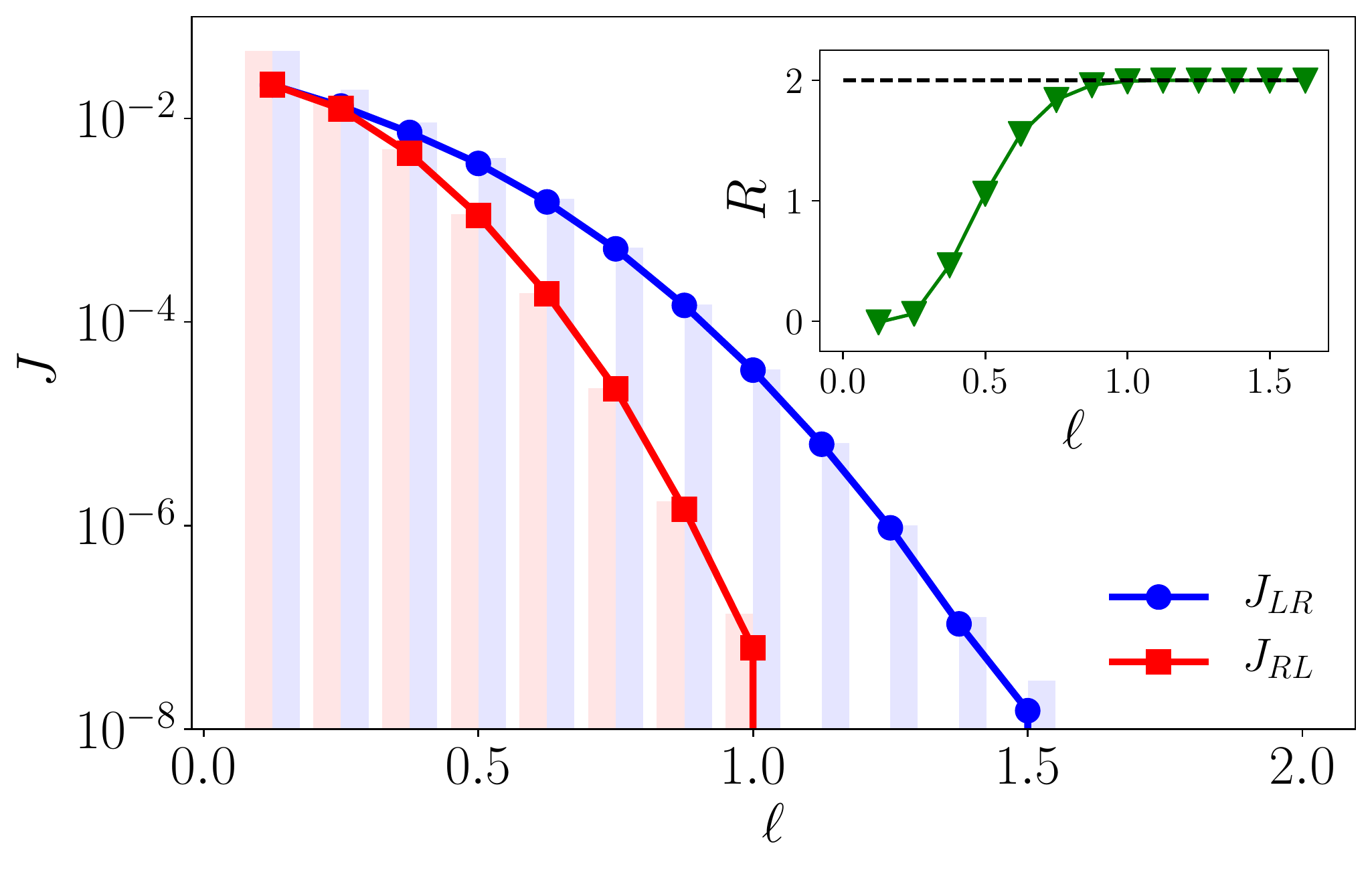}	
	\caption{ Currents for a harmonic system with two particles ($N=2$) as a function of the  width $\ell$ of the {\bf infinite-square-well interface}.     The   inset shows the rectification factor and the dashed horizontal line refers to the {\bf perfect diode} effect, nearly attained for $\ell \gtrsim 1$.
	The histogram of $\rho \Delta T$, where $\rho$ is the the number of stretches per unit time, is also shown in the main plot.  
	$T_m = 0.05$ and $\Delta_{rel} = 1$. In this case, since $N=2$, we set $k_L = k_R = 0$. 
 }
	\label{fig:infinite}
\end{figure}

In order to provide an intuitive understanding of the limit $\mu\to\infty$,  we investigate heat conduction through the infinite square well, with a half-width $\ell$, at the interface. The impact of the length $\ell$ on the currents and on the rectification factor can be visualized in Fig.~\ref{fig:infinite}. 
To facilitate the comparison with heuristic considerations, we chose the simplest case of two particles ($N=2$) subject to  harmonic on-site potentials.

 Energy transfer across the interface occurs exclusively by means of collisions with the potential walls, specifically when the relative displacement reaches $r = x_2-x_1-a = \pm \ell$. Beyond these instantaneous collisions, the interface particles, and hence the chains, move completely independently of each other.  
Since we examine the system under low-temperature conditions,  the rate of collisions, $\rho$, which is defined as the number of collisions divided by the observation time, is relatively low. Then, the inter-collision time is sufficiently long to allow the particles to nearly reach thermal equilibrium with their respective heat baths. 
During a single collision, the particles, having equal masses, exchange their kinetic energies, resulting in $\Delta K = m \langle v_1^2 \rangle - m \langle v_2^2 \rangle \approx \Delta T$.
Therefore, the current can be approximated as the product of the collision rate $\rho$ and the kinetic energy exchange, given by $J \approx \rho \Delta T$. The effectiveness of this approximation is demonstrated in Fig. \ref{fig:infinite}.

To grasp the influence of asymmetry, we examine a scenario analogous to that depicted in  Fig.~\ref{fig:infinite}, namely, a system with $N=2$ particles experiencing harmonic on-site potentials, where $A_L=1$, $A_R=5$, $T_m=0.05$, $\Delta_{rel}=\pm 1$,  and $k_L=k_R = 0$.  
Since we are considering the rare-collision regime, we can assume that each particle reaches thermal equilibrium with its corresponding bath, 
 in such case the variance of the relative motion is $\langle r^2 \rangle = T_L/A_L + T_R/A_R$. 
  Notably, this variance exhibits clear asymmetry upon exchange of thermal bath temperatures. Specifically, for the case of $T_L > T_R$, the variance is 0.08, whereas for $T_R > T_L$, the variance is 0.04. 
As a consequence, 
the probability of observing $|r| > \ell$ is approximately 700 times higher 
in the former compared to the latter case. 
This estimate is in agreement with our numerical outcomes, 
where the typical time interval between two successive collisions is also roughly 700 times longer when comparing $T_L > T_R$ with $T_L < T_R$.  
Such significant disparity is explained by the fact that both probabilities are dominated by rare events. As a result, even a slight alteration in the variance can lead to an enormous consequence.
We have observed that our theoretical estimate remains valid for  higher values of $T_m$ (not shown).

When $\ell \to 0$, the relative coordinate tends to be confined near zero, therefore, fluctuations, and hence the asymmetry, are suppressed. In this limit, the dominant motion is that of the center of mass, and the currents in each direction coincide.  
In the opposite limit of large $\ell$, at low enough temperature, a maximal efficiency close to $R=2$ (perfect diode) is attained when $\mu \to \infty$. For instance,  in the case of Fig. \ref{fig:infinite}, $\ell=1$ is enough to achieve the perfect diode within 0.5\% tolerance ($R=1.992$).
In what follows we will set $\ell=1$.

\subsection*{Effect of other parameters}
\label{sec:other}

 In this section, we inspect the effects of $\Delta_{rel}$, $k_\mu$ and $N$, at low $T_m$. As paradigmatic on-site potential, we use the $\phi^4$ model.

\begin{figure}[h!]
	\centering
	\includegraphics[width=0.45\textwidth]{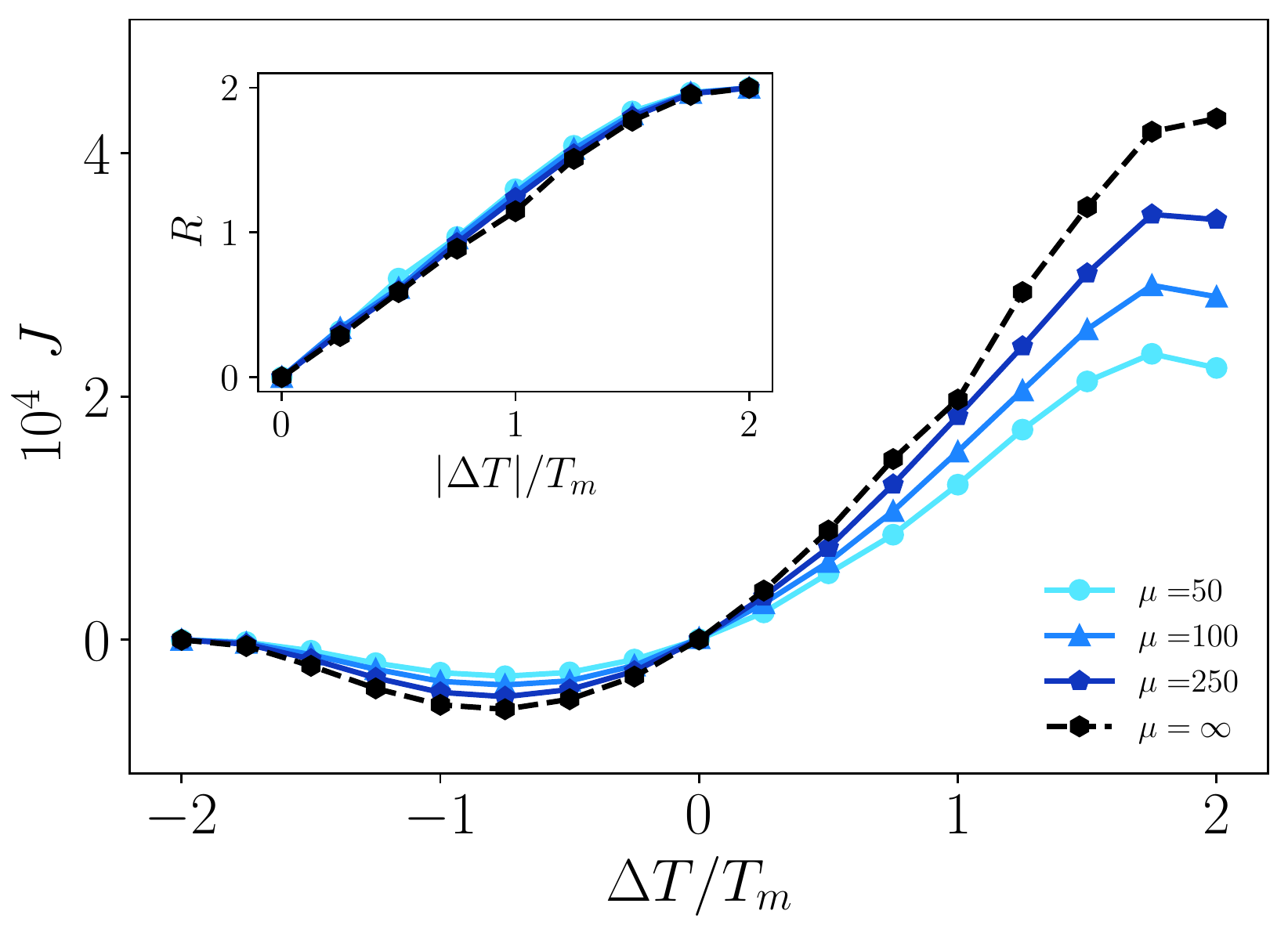}
	\caption{  
 Heat flux vs. $\Delta_{rel} \equiv \Delta T/T_m$ for  different values of $\mu$, setting $N=20$, $k_\mu=0.5$, $T_m=0.05$ and the $\phi^4$ model. The inset shows the corresponding rectification factor. 
 }
	\label{fig:dT}
\end{figure}

\begin{figure}[h!]
	\centering
	\includegraphics[width=0.45\textwidth]{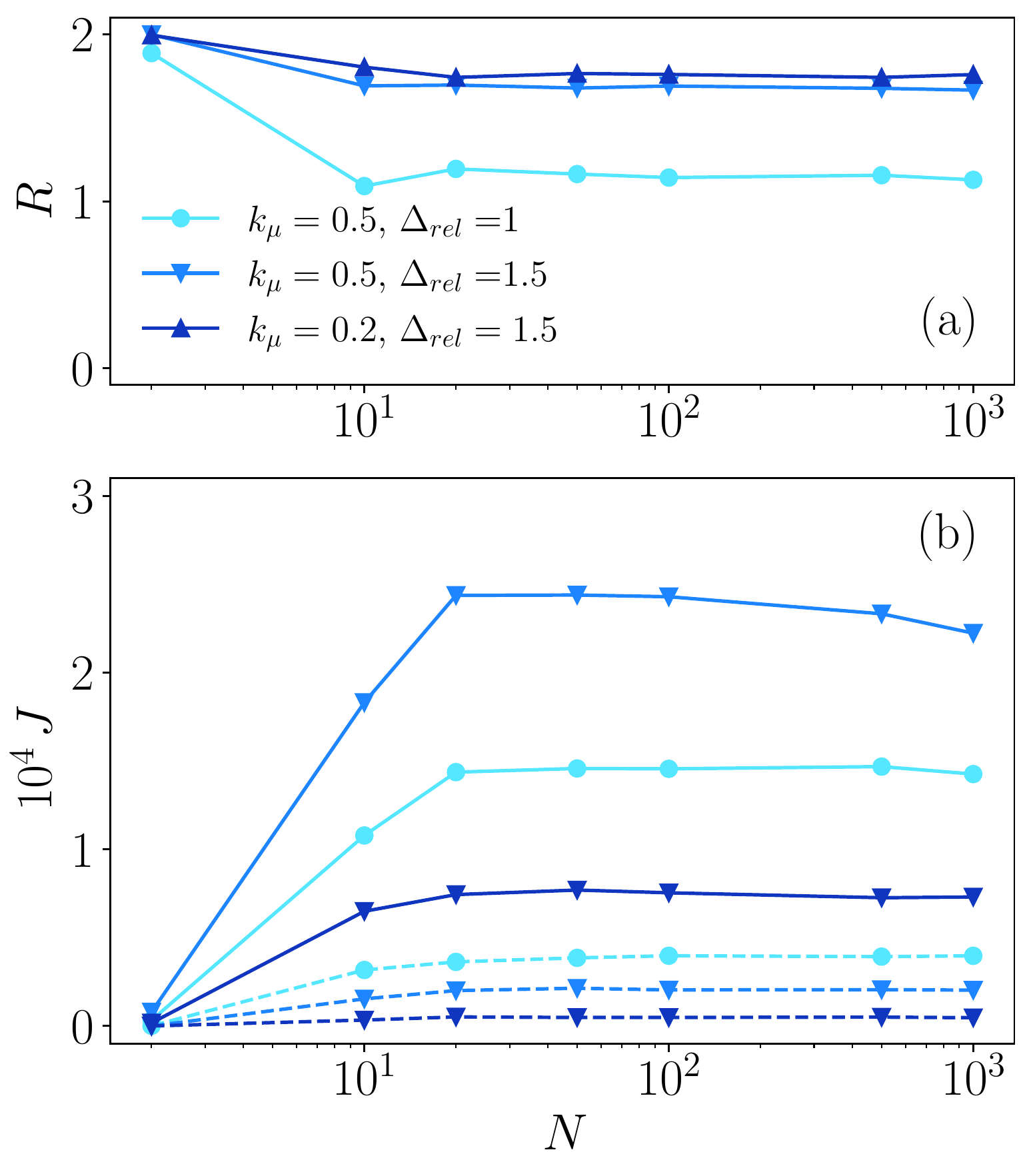}\\[-3mm]
	\caption{  
Size effects. Rectification and corresponding heat fluxes vs. $N$, for the $\phi^4$ on-site potential, and interface with $\mu=10$, in the cases:    
$(k_\mu,\Delta_{rel})=$ (0.5,1.0),(0.5,1.5), (0.2,1.5), with $T_m=0.05$.
}
	\label{fig:N}
\end{figure}

{\bf Relative temperature difference $\Delta_{rel}$. } 
The effect of $\Delta_{rel}$, at $T_m=0.05$ and $k_\mu=0.5$, 
is illustrated in Fig.~\ref{fig:dT} for the case of $N=20$ particles subject to    $\phi^4$ on-site potential, considering  different values of the interface exponent $\mu$  approaching the infinite-$\mu$ limit. 
In the inset, we observe that the corresponding rectification is enhanced when increasing $|\Delta_{rel}|$, approaching the perfect diode when the relative temperature difference is maximal.  
The effect of $|\Delta_{rel}|$ is also illustrated in  Fig.~\ref{fig:N}, for $\mu=10$ and different values of the chain length $N$.

{\bf Stiffness constant $k_\mu$. } 
Diminishing $k_\mu$ can enhance rectification, as can be observed in Fig. ~\ref{fig:N}. 
for $\mu=10$. 
In fact, according to the definition of the potential Eq.~(\ref{eq:Vmu}), 
diminishing $k_\mu$ is directly related to increasing $\ell$, which favors rectification as discussed in Sec.~\ref{sec:infinite}. 

{\bf System size $N$.} 
The effect of system size is illustrated in Fig.~\ref{fig:N}, where we have plotted the rectification factor as a function of   $N$, for $\mu=10$, using different values of $(k_\mu,\Delta_{rel})$.
Increasing system size from $N=2$, we observe an initial decrease of the rectification, and further decay is expected, as the system size is known to reduce rectification~\cite{casati-diode}.  
At the same time, the figure illustrates the consequences of diminishing $k_\mu$, and augmenting $\Delta_{rel}$, both contributing to improve rectification.

\section{Conclusions}
\label{sec:final}
 
We studied the diode effect connecting Langevin thermal baths at the ends of one-dimensional chains formed by two different segments, with interfacial particles coupled through the nonlinear force derived from Eq.~(\ref{eq:Vmu}).  
This type of nonlinear response means a deformation-dependent stiffness 
$\bar{k}(x)=k_\mu |x|^{\mu-2}$, where $x$ is the deformation, which is constant in the harmonic case $\mu=2$, but increases ($\mu>2$) or decreases ($1\le\mu<2$) with the deformation otherwise. 
We inspected the full range of $\mu$, at different temperatures, but focused on the rectification performance near the infinite-square well limit ($\mu\to\infty$) under low-temperature conditions. 

We observed that,  for all the on-site potentials considered,  
 thermal rectification is maximal at a value of  $\mu$ that increases when $T_m$ is decreased. 
 This suggests that to optimize rectification at low temperatures, the effective stiffness of the interfacial restoring force must have a rapidly increasing dependency on the deformation.
Moreover,
 the perfect-diode effect can be approached when $\mu \to \infty$, within a desired accuracy,  not only  lowering $T_m$, but also  increasing  $\Delta_{rel}$, or  
diminishing $k_\mu$. 
We highlight that the temperature scale is directly related to the choice of parameters. However, the behavior observed in the infinite-square well limit under low-temperature conditions is still expected to hold.

Beyond low-temperature conditions, the optimal rectification can be achieved at a finite value of $\mu$. 
High values of $R$ can also emerge in this regime, but the optimization conditions 
are dominated not only by the interface characterized by $\mu$ but also by the type of on-site potential, and hence vary depending on the values of system and bath parameters. 
Particularly, depending on the on-site potential, 
the optimal $\mu$ can correspond  to a stiffness with superlinear ($\mu>2$) or sublinear ($\mu<2$) dependency on the deformation, 
as observed from the position of the maxima in Figs.~\ref{fig:R20-fi4}, \ref{fig:R20-harmonic}, and \ref{fig:R20-FK}. 
Regarding the interplay of the interface with the on-site potential, let us note that for harmonic segments, the interfacial nonlinearity is enough to produce temperature dependency and allows to optimize rectification.   
Remarkable features are observed for the FK chains, where  
the exponent $\mu$ has the additional role of inducing the inversion of the preferential direction, as seen in Fig.~\ref{fig:R20-FK}.

We adopted a particular form of producing asymmetry, namely, we used chains with a unique type of on-site potential, with different values of parameters for each segment. But, of course, there is a plethora of other possibilities to investigate, such as different on-site potentials for each segment, identical potentials but different  masses~\cite{defaveri2021}, asymmetric potentials~\cite{muga2017}, among many other combinations, which can motivate future continuations. 
It could be also interesting to perform a systematic study using different bath models and coupling schemes, under the light of ref.~\cite{Li2010}.


\begin{thebibliography}{99}

\bibitem{starr36}
C. Starr,
{\it The Copper Oxide Rectifier},  
Physics, 7, 15 (1936).

\bibitem{review}
M.Y. Wong, C.Y. Tso , T.C. Ho, H.H. Lee,  
{\it  
A review of state of the art thermal diodes and their potential applications}. 
International Journal of Heat and Mass Transfer 164, 120607 (2021).
 
\bibitem{terraneo2002}
M. Terraneo, M. Peyrard, G. Casatti,
 {\it Controlling the
  energy flow in nonlinear lattices: A model for a thermal rectifier},
Phys. Rev. Lett.  88,  4 (2002).

\bibitem{casati-diode}
B. Li, L. Wang, G. Casati, 
{\it Thermal diode: Rectification of heat flux}, 
Phys. Rev. Lett.  93, 184301 (2004).

  

\bibitem{efficiency1}
E. Pereira, R.R. \'Avila, 
{\it Increasing thermal rectification: Effects
of long-range interactions},  
Phys. Rev. E, 88, 032139 (2013).

\bibitem{efficiency2}
S. Chen, E. Pereira, G. Casati, 
{\it Ingredients for an efficient
thermal diode},  EPL   111(3), 30004 (2015).
 
 \bibitem{efficiency3}
S. Chen, D. Donadio, G. Benenti, G. Casati, 
{\it Efficient thermal diode with ballistic spacer},  Phys. Rev. E, 97, 030101 (2018).

  \bibitem{changSolidstate2006} 
C.W. Chang, D. Okawa,  A. Majumdar, A. Zettl,
{\it Solid-state thermal rectifier}, 
Science, 314(5802), 1121–1124 (2006).
  
\bibitem{kobayashiOxideThermal2009}%
W. Kobayashi, Y. Teraoka, I. Terasaki,  
{\it  An oxide thermal rectifier}, 
Applied Physics Letters 95, 171905 (2009).

 \bibitem{tianNovelSolidstate2012}%
 H. Tian, D. Xie, Y. Yang, T. Ren, G. Zhang, Y. Wang, C. Zhou, P. Peng, L. Wang, L. Liu, {\it  A novel solid-state thermal rectifier based on reduced graphene oxide},
 Sci. Rep. 2, 523 (2012). 
 
 \bibitem{martinez-perezRectificationElectronic2015}%
 M. J. Mart\'inez-P\'erez, A. Fornieri, F. Giazotto, {\it Rectification of electronic heat current by a hybrid thermal
diode},  Nature Nanotechnology 10, 303 (2015). 

 \bibitem{tsoSolidstateThermal2016}%
C. Tso, C. Y.  Chao, 
{\it  Solid-state thermal diode with
shape memory alloys},  Int. J.  of Heat
and Mass Transfer 93, 605 (2016).

  \bibitem{shresthaDualmodeSolidstate2020}%
  R. Shrestha, Y. Luan, X.  Luo, S. Shin, T.  Zhang,  
P. Smith,   W. Gong, M.  Bockstaller, T. Luo,  R. Chen, K.  Hippalgaonkar, S.  Shen,  
{\it  Dual-mode solid-state
thermal rectification}, 
Nature Comm. 11, 4346 (2020).
  
 \bibitem{kasprzakHighTemperature2020}%
 M. Kasprzak, M. Sledzinska, K. Zaleski, I. Iatsunskyi, F.  Alzina, S. Volz, C.M.  Sotomayor Torres, B. Graczykowski, 
 {\it  High temperature silicon thermal
diode and switch},  Nano Energy 78, 105261 (2020).
 
\bibitem{defaveri2021}
  L. Defaveri,  C. Anteneodo,  
 {\it Analytical results for a minimalist thermal diode}, 
Phys. Rev. E 104, 014106 (2021).
 
 \bibitem{muga2021}
M. A. Sim\'on, A. Ala\~na, M. Pons, A. Ruiz-Garcia, and J.
G. Muga. 
{\it Heat rectification with a minimal model of two
harmonic oscillators}, Phys. Rev. E 103, 012134 (2021).

\bibitem{muga2023}
J. Navarro, J. G. Muga, and M. Pons, 
{\it Heat rectification, heat fluxes, and spectral matching},  arXiv:2302.13874v1.

\bibitem{Tdependence}
E. Pereira
{\it Requisite ingredients for thermal rectification}
Phys. Rev. E 96, 012114 (2017). 

\bibitem{bastida}
M. Romero-Bastida, J.O.M. Pe\~na, J.M. L\'opez, 
{\it  Thermal rectification in mass-graded next-nearest-neighbor Fermi-Pasta-Ulam lattices}, 
Phys. Rev. E 95, 032146 (2017).

\bibitem{benenti}
G. Benenti, G. Casati, C. Mejía-Monasterio, M. Peyrard
in "Thermal Transport in Low Dimensions: From Statistical Physics to Nanoscale Heat Transfer",S. Lepri (editor),
(Springer International Publishing, 2016).

\bibitem{muga2017}
M. Pons, Y. Y. Cui, A. Ruschhaupt, M. A. Sim\'on and J. G. Muga,
{\it Local rectification of heat flux}, 
EPL 119 64001 (2017).

 \bibitem{alexander2020}
T. J. Alexander, 
{\it  High-heat-flux rectification due to a localized thermal diode}, 
Phys. Rev. E 101, 062122 (2020).


\bibitem{strain1}
K. Ren, H.  Qin, H.  Liu, Y.  Chen, X. Liu, G. Zhang, 
{\it Manipulating Interfacial Thermal conduction of 2D Janus heterostructure via a thermo-mechanical coupling}, 
Adv. Functional Materials 32, 2110846 (2022). 

\bibitem{strain2}
F.Liu, Y.K. Gong, R, Zou, H. Ning, N. Hu, Y. Liu, L. Wu, F. Mo, S. Fu, C. Yan, 
{\it Strain effects on the interfacial thermal conductance of graphene/h-BN heterostructure}, 
Nano Materials Science 4, 227 (2022).

\bibitem{interface}
J. Wang and Z. Zheng, 
{\it Heat conduction and reversed thermal diode: The interface effect},
Phys. Rev. E 81, 011114 (2010).
  
\bibitem{BaowenLi2005}
 B. Li, J. Lan, and L. Wang, 
{\it Interface thermal resistance between dissimilar anharmonic lattices}, 
Phys. Rev. Lett. 95, 104302 (2005).

\bibitem{BaowenLi2011}
L. Zhang, P. Keblinski, J.-S. Wang, and B. Li.
{\it Interfacial thermal transport in atomic junctions}, 
Phys. Rev. B 83, 064303 (2011).

\bibitem{Hu2006} 
B. Hu, L. Yang, Y. Zhang., Phys. Rev. Lett.  97. 
124302 (2006). 

\bibitem{hanggi} 
 S. Liu, J. Liu, P. Hanggi, C. Wu, and B. Li, 
 {\it Triggering waves in nonlinear lattices: Quest for anharmonic phonons
and corresponding mean-free paths}, Phys. Rev. B 90, 174304 (2014).

\bibitem{defaveri2019}
E. H. Colombo ,
L. A. C. A. Defaveri,  C. Anteneodo, 
Phys. Rev. E 100, 032118 (2019).

\bibitem{dhar2019} 
A. Dhar, A. Kundu, S. N. Majumdar, S. Sabhapandit, and G.
Schehr, Phys. Rev. E 99, 032132 (2019).
 
\bibitem{schmelcher2018}
[25] P. Schmelcher, Phys. Rev. E 98, 022222 (2018).

\bibitem{meiners2000}
J.-C. Meiners and S. R. Quake, Phys. Rev. Lett. 84, 5014 (2000).

\bibitem{NarayanRamaswamy2002}
O. Narayan, S. Ramaswamy, 
Phys.  Rev.  Lett.  89, 200601 (2002).

\bibitem{ReviewLepriLiviPoliti2003}
S. Lepri, R. Livi,  A. Politi, 
{Phys. Rep.}  377, 1  (2003).

\bibitem{ReviewDhar2008}
A. Dhar, 
{\it Heat transport in low-dimensional systems}, 
Adv. in Physics   57, 457  (2008).


\bibitem{aokiEnergyTransport2006}%
K. Aoki, J. Lukkarinen, H. Spohn, 
{\it  Energy transport in
weakly anharmonic chains}, 
Journal of Statistical
Physics 124, 1105 (2006).

\bibitem{savinHeatConduction2003}%
 A. V. Savin,  Gendelman, 
 {\it  Heat conduction in one-
dimensional lattices with on-site potential},  Phys. Rev. E 67, 041205 (2003).
 
\bibitem{huHeatConduction2000}%
 B. Hu,  B. Li, H. Zhao, 
 {\it  Heat conduction in one-
dimensional nonintegrable systems},  
Phys. Rev. E 61, 3828 (2000).

\bibitem{lefevereNormalHeat2006}%
R. Lefevere, A. Schenkel, {\it  Normal heat conductivity in
a strongly pinned chain of anharmonic oscillators},  
JSTAT  2006, L02001 (2006).

\bibitem{liParameterdependentThermal2007} 
N. Li and B. Li, 
{\it Parameter-dependent thermal conductivity of one-dimensional $\phi^4$ lattice}, 
Phys. Rev. E 76, 011108( 2007).

\bibitem{huHeatConduction1998}%
B. Hu, B. Li, H. Zhao, 
{\it Heat conduction in one-dimensional chains}, 
Phys.  Rev.  E 57, 2992 (1998).

\bibitem{Gillan1985i}  
M.J. Gillan, 
{\it Transport in the Frenkel-Kontorova model. I. Diffusion and single-particle motion}, 
J. Phys. C 18, 4485 (1985).
 
\bibitem{Gillan1985ii}  
M.J. Gillan, R.W. Holloway,   
{\it Transport in the Frenkel-Kontorova model. II. The diffusion coefficient}, 
J. Phys. C 18, 4903 (1985).

\bibitem{Gillan1985iii}  
M.J. Gillan, R.W. Holloway,   
{\it Transport in the Frenkel-Kontorova model. III. Thermal conductivity}, 
J. Phys. C 18, 5705 (1985).

\bibitem{Penland2006}
J. A. Hansen and C. Penland,
\textit{Efficient Approximate Techniques for Integrating Stochastic Differential Equations},
\href{https://doi.org/10.1175/MWR3192.1}{Monthly Weather Review {\bf 134}, 3006–3014} (2006)

\bibitem{Li2010}
J. Chen, G. Zhang, B. Li,
{\it Molecular Dynamics Simulations of Heat Conduction in Nanostructures: Effect of Heat Bath},
Journal of the Physical Society of Japan, 79, 074604 (2010).

\end{thebibliography}

{\bf Acknowledgements:} 
C.A. acknowledges partial financial support obtained from CNPq (311435/2020-3), and FAPERJ (CNE E-26/201.109/2021) in Brazil.


\appendix
 
 \setcounter{figure}{0}
  \renewcommand{\thefigure}{A\arabic{figure}}
\setcounter{subsection}{0}
\section{Rectification for other on-site potentials}

\begin{figure}[h!]
    \includegraphics[scale=0.45]{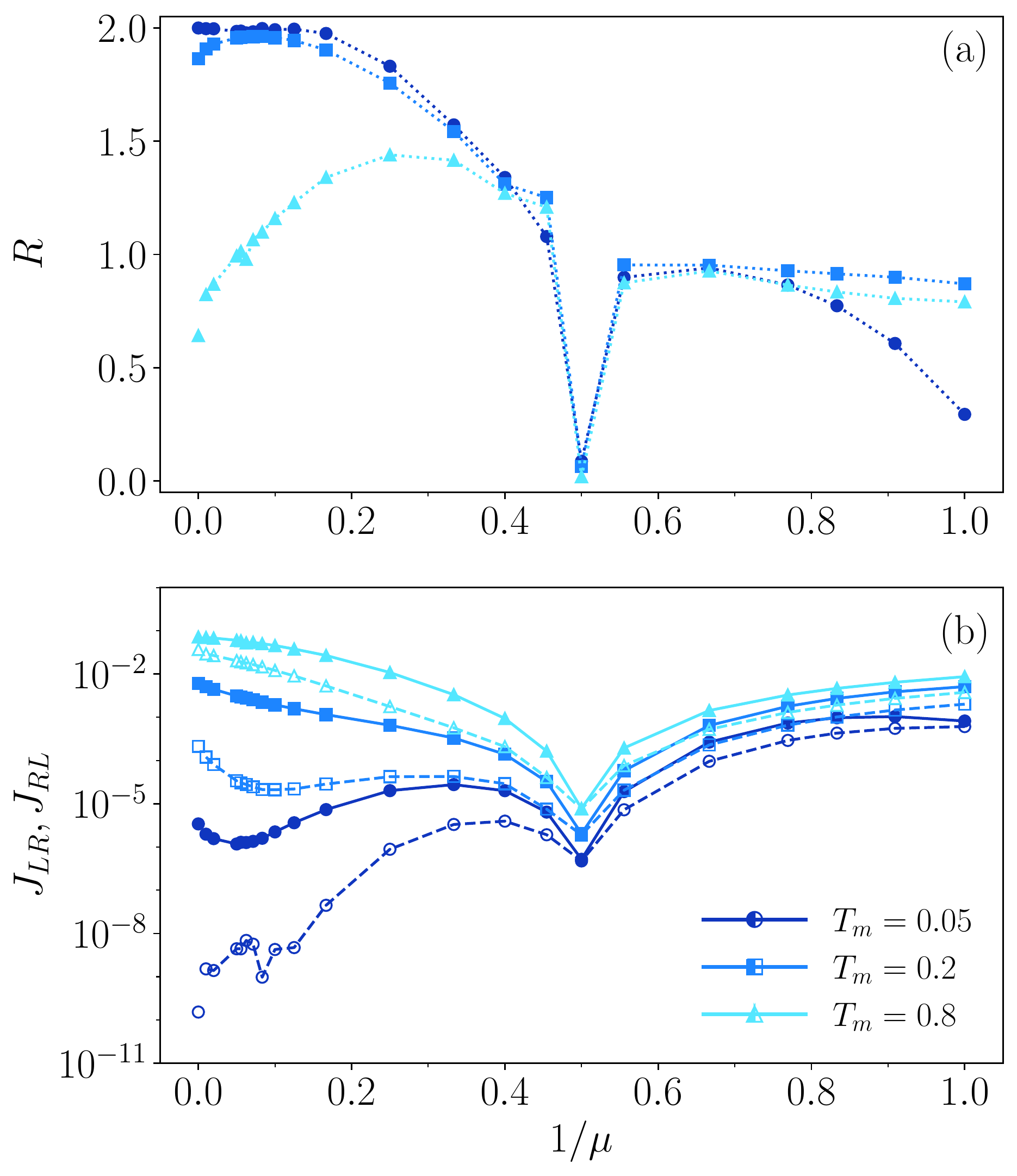}
    \caption{
 (a) Rectification factor $R$ and (b) corresponding heat currents $J_{LR}$ and $J_{RL}$ vs. $1/\mu$, 
   for {\bf harmonic on-site} potential, and different values of  $T_m$.
Other details and values of the parameters are the same used in Fig.~\ref{fig:R20-fi4}.  
}
    \label{fig:R20-harmonic}
\end{figure}

\begin{figure}[h!]
    \includegraphics[scale=0.45]{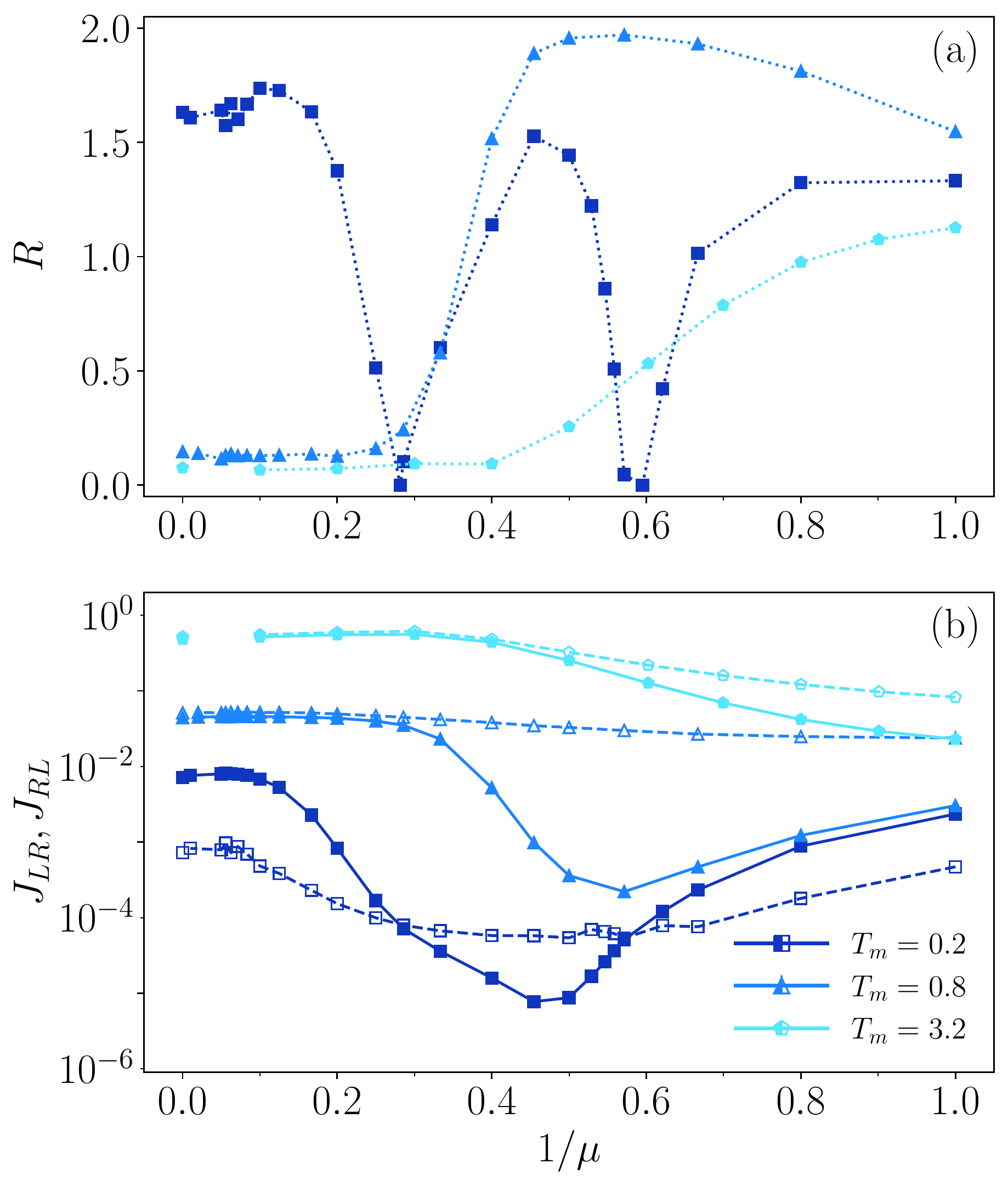}
    \caption{
 (a) Rectification factor $R$ and (b) corresponding heat currents $J_{LR}$ and $J_{RL}$ vs. $1/\mu$, 
    for {\bf FK on-site} potential,
and different values of  $T_m$. 
Other details and values of the parameters are the same used in Fig.~\ref{fig:R20-fi4}.
$R=0$ was attributed to the values of $\mu$ at which inversion of the preferential direction occurs. 
}
    \label{fig:R20-FK}
\end{figure}

The rectification and corresponding currents, for chains with $N=20$ particles,  subject to the harmonic and FK on-site potentials, are respectively shown in Figs.~\ref{fig:R20-harmonic} and 
\ref{fig:R20-FK}, for different temperatures $T_m$. In both cases, 
for small enough $1/\mu$, we observe a behavior similar to that shown for the $\phi^4$ potential in  Fig.~\ref{fig:R20-fi4}. 
Indeed,  when $T_m$ decreases, the rectification $R$ attains a maximal level, at values of $\mu$ that approach the infinite-square well limit.
For the infinite-square-well, the rectification $R$ vs. $T_m$ was shown in the inset of Fig.~\ref{fig:R20-fi4}, for all the three studied on-site potentials.

In contrast, 
for not too large $\mu$, there are details specific to each potential.

In the harmonic case (Fig.~\ref{fig:R20-harmonic}), $R$ abruptly vanishes (within the standard deviation) at $\mu=2$, where the system becomes fully harmonic and hence the currents in both directions identical.

The FK potential, which has the distinctive feature of being bounded, presents a more complex behavior (Fig.~\ref{fig:R20-FK}), where more than one value of $\mu$ for inversion of the preferential direction and more than one maximal value of $R$ can occur.  
It is interesting to note that at fixed $T_m$, current inversion can be switched by $\mu$ in the FK chain. 
Another remarkable feature is that a high level of rectification can be attained at relatively high temperatures, for the $\phi^4$ and FK chains, as can be observed in Figs.~\ref{fig:R20-fi4} and \ref{fig:R20-FK}. 
 
In the $\phi^4$ model, when the temperature increases, the maximal efficiency shifts towards the triangular potential ($\mu=1$), but such effect is not observed for the harmonic chains, and certainly not for the FK chains, where the bounded on-site potential will become irrelevant at high temperatures, with the consequent decrease of the diode effect.  

\end{document}